\documentclass[10pt,letterpaper,twocolumn]{article} 

\usepackage{amssymb}
\usepackage{amsmath,mathrsfs,amsfonts}
\usepackage{ol2}
\usepackage[draft]{hyperref}
\usepackage[caption=false]{subfig}

\begin{document}

\twocolumn[ 

\title{Unidirectional and Wavelength Selective Photonic Sphere-Array Nanoantennas}


\author{Yang G. Liu,$^{1,2}$ Wallace C.H. Choy,$^{2}$ Wei E.I. Sha,$^{2,*}$ and Weng Cho Chew$^{2,3,*}$}

\address{
$^1$Institute of Applied Physics and Computational Mathematics, Fenghao East Road, Beijing, China.\\
$^2$Department of Electrical and Electronic Engineering, the
University of Hong Kong, Pokfulam Road, Hong Kong.\\
$^3$Department of Electrical and Computer Engineering, University of Illinois, Urbana-Champaign, USA.\\
$^*$Corresponding authors: wsha@eee.hku.hk (W.E.I. Sha); wcchew@hku.hk (W.C. Chew).
}

\begin{abstract}
We design a photonic sphere-array nanoantenna (NA) exhibiting both strong directionality and wavelength selectivity. Although the geometric configuration of the photonic NA resembles a plasmonic Yagi-Uda NA, it has different working principles, and most importantly, reduces the inherent metallic loss from plasmonic elements. For any selected optical wavelength, a sharp Fano-resonance by the reflector is tunable to overlap spectrally with a wider dipole resonance by the sphere-chain director leading to the high directionality. The work provides design principles for directional and selective photonic NAs, which is particularly useful for photon detection and spontaneous emission manipulation.
\end{abstract}


] 

Optical nanoantennas (NAs) \cite{Ref1,Ref2,Ref3,Ref4,Ref5} have been a hot research topic due to their abilities to control light in a subwavelength scale. While many studies have focused on the near-field enhancement in the past several years, directional far-field properties of NAs have attracted more and more attentions recently. Developing a directional NA to redirect the emission from an ensemble of atoms or molecules with random dipole orientations is particularly important to photon detection and sensing, spectroscopy and microscopy, and spontaneous emission manipulation \cite{Ref3,Ref4,Ref6,Ref7,Ref8,Ref9,Ref10,Ref11}. Although various plasmonic NAs have been reported in the literature to realize the directional functionality, they suffer from a fundamental limit due to intrinsic metallic loss.

In this letter, we design a photonic sphere-array NA to reduce the ohmic loss and maintain other useful functionalities of plasmonic NAs. The geometric configuration of the photonic NA resembles a plasmonic Yagi-Uda NA \cite{Ref12,Ref13,Ref14,Ref15}, which comprises a reflector with a single sphere and a director with a sphere chain. However, the working principle of the photonic NA distinguishes from its plasmonic counterpart. A narrow Fano-resonance by the reflector is manipulated to overlap spectrally with a wider dipole resonance by the director giving rise to both high directionality and wavelength selectivity. In particular, the wavelength selectivity by the narrow and sharp Fano-resonance is required to resolve and match the vibrational modes of the target molecule. As an accurate and fast solution to Maxwell's equations, the T-matrix method \cite{Ref16,Ref17,Ref18} with a Hertzian dipole source is adopted to analyze and optimize the far-field response of the photonic sphere-array NA.

\begin{figure}[!tbc]
\centering
\includegraphics[width=2.5in]{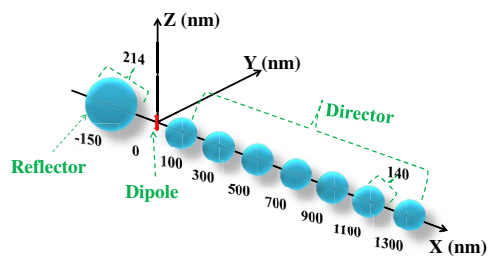}
\caption{The schematic design for an optimized photonic sphere-array NA at the selected wavelength of $603$ nm.}
\label{fig1}
\end{figure}

\begin{figure}[!tbc]
\centering
\includegraphics[width=2.8in]{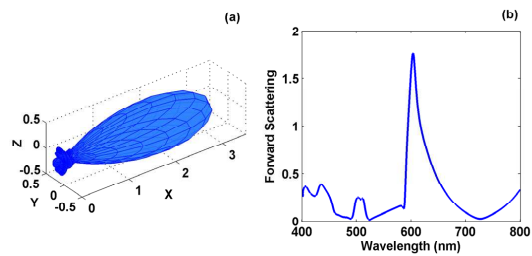}
\caption{(a) The radiation pattern of the optimized photonic NA at a selected wavelength of $603$ nm; (b) The forward scattering intensity (along the $x$ direction) of the photonic NA as a function of the wavelength.}
\label{fig2}
\end{figure}

Fig. \ref{fig1} shows an optimized design of the silicon sphere-array NA. Interestingly, the geometrical configuration of the reflector and director are tunable to a selected wavelength with a high directionality as demonstrated in Fig. \ref{fig2}. Directivity measures the power density the antenna radiates in the direction of its strongest emission, versus the power density radiated by an ideal isotropic radiator (which emits uniformly in all directions) radiating the same total power. The directivity of the optimized NA is 15.7 compared to 1.5 for a short dipole antenna. We will study the working principle of the reflector and director separately. A $z$-polarized Hertzian dipole emitter is used to excite the NA.

\begin{figure}[!tbc]
\centering
\includegraphics[width=2.8in]{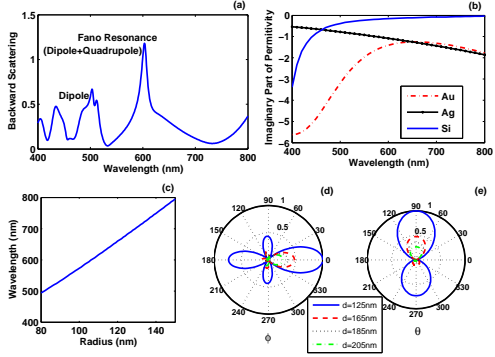}
\caption{(a) The backward scattering intensity of a silicon nanosphere (with the radius of $107$ nm) as a function of the wavelength; (b) The imaginary parts of the relative permittivities of the gold, silver, and silicon; (c) The tunable Fano resonance by varying the sphere radius; (d) The radiation patterns at the $xoy$ plane after modifying the separation $d$ between the dipole emitter and the center of the sphere; (e) The radiation patterns at the $zox$ plane.}
\label{fig3}
\end{figure}

Fig. \ref{fig3}(a) shows the backward scattering intensity (along the $x$ direction) of a photonic nanosphere with the radius of $107$ nm. The backward scattering amplitude can be calculated by the T-matrix method. Each summation term of the spherical harmonics can be viewed as an angular momentum or multipole channel with different resonance linewidths strongly depending on the dispersive and dissipative properties of the silicon material as depicted in Fig. \ref{fig3}(b). In comparison with metal, silicon can reduce the ohmic loss. The constructive and destructive interferences between the broad dipole eigenmode (first term of the spherical harmonics) and the narrow quadrupole eigenmode (second term of the spherical harmonics) induce an extraordinary Fano resonance with an asymmetric spectral line and significant directionality. As shown in Fig. \ref{fig3}(a), the Fano resonance peak at 603 nm is significantly sharper than other dipole resonance peaks before 550 nm. Importantly, we reveal that the Fano resonance can be supported by a single photonic nanosphere in contrast with previous works where the Fano resonance occurs in a plasmonic nanosphere or nanodisk \cite{Ref19,Ref20,Ref21,Ref22,Ref23}. Fig. \ref{fig3} (c) demonstrates that the Fano resonance is highly tunable by engineering the nanosphere radius. It is worth mentioning that both plasmonic and photonic nanospheres supporting the dipole resonance alone with small sizes cannot produce any Fano resonances (accurate wavelength selectivity) in the visible light range. Figs. \ref{fig3}(d) and (e) show the radiation patterns of the nanosphere respectively at the $xoy$ and $zox$ planes. The separation between the nanosphere and the dipole emitter plays a key role in the far-field response of the reflector. At a shorter separation, one observes a gradually increasing the backward scattering spectra. However, the directionality becomes insignificant for the ignorable field retardation ($d = 125$ nm). The strong near-field interplay between the silicon nanosphere and the emitter decreases the directionality with unwanted side lobes. Moreover, the strong coupling induces a non-radiative loss that is responsible for the quenched fluorescence from the emitters \cite{Ref9}. Hence, a specific degree of field retardation is essential for the Fano-resonance-boosted directionality.

\begin{figure*}[!tbc]
\centering
\includegraphics[width=3.8in]{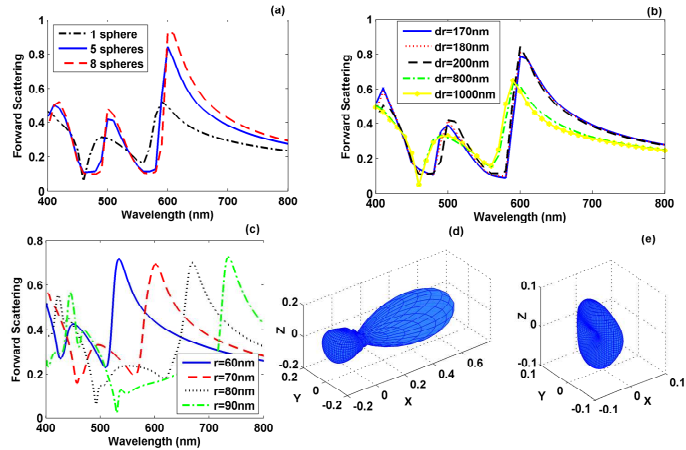}
\caption{(a) The forward scattering intensity for three directors comprising $1$, $5$ and $8$ silicon nanospheres, respectively. The radius of each sphere is $70$ nm; (b) The forward scattering intensity for the director (involving $5$ silicon nanospheres) as a function of the periodicity of the sphere chain; (c) The forward scattering intensity of the director as a function of the sphere radius; (d) The radiation pattern of the director working at the dipole resonance peak ($603$ nm); (e) The radiation pattern of the director at off-resonance ($565$ nm).}
\label{fig4}
\end{figure*}

A director comprising a set of interacting photonic nanospheres also strongly affects the directionality of the optical NA. Fig. \ref{fig4}(a) illustrates the forward scattering intensity (along the $x$ direction) of three directors with $1$, $5$ and $8$ silicon nanospheres, respectively. The nanospheres of the director have the same radius $r$ and are arranged in a chain along the $x$ axis with a fixed periodicity $d_r$, i.e., the distance between centers of two adjacent nanospheres. Here we set $r=70$ nm and $d_r=200$ nm. The separation between the center of the first sphere and the emitter is assumed to be $d_s=120$ nm. From Fig. \ref{fig4}(a), we can see that the wavelength corresponding to the maximum forward scattering is not sensitive to the number of spheres. The forward scattering peaks of multiple spheres almost coincide with the dipole resonance peak of a single photonic nanosphere with a little peak shifting. Intriguingly, the radiation pattern alters drastically from on-resonance ($603$ nm) as presented in Fig. \ref{fig4}(d) to off-resonance ($565$ nm) as presented in Fig. \ref{fig4}(e). In Fig. \ref{fig4}(c), the forward scattering peak of the director involving $5$ spheres is highly tunable by manipulating the sphere radius governing the dipole resonance wavelength. Fig. \ref{fig4}(b) shows the perturbed forward scattering peak under the influence of the dipole-dipole coupling with the modified periodicity $d_r$. For a large periodicity, the nanospheres can be regarded as independent scatterers, and thus the beamforming of the director is essentially a far-field interference or antenna synthesis problem. For a small periodicity, the dipole-dipole interaction shifts the resonance peaks. However, the secondary dipole associated with each photonic element induced by other adjacent ones is expected to be weak due to the $z$-polarized emitter and should be weaker than that in a plasmonic system.

For any selected optical wavelength, the pronounced Fano resonance by the reflector is tunable to overlap spectrally with a wider dipole resonance by the director leading to high directionality and selectivity of the photonic NA as shown in Figs. \ref{fig2}(a) and (b), respectively. The realization of the spectral overlap is trivial owing to the highly tunable properties of both the reflector and director discussed above. Taking the selected wavelength of $603$ nm as an example, the photonic NA has a good wavelength selectivity with a bandwidth that is wider than that of the reflector as shown in Fig. \ref{fig3}(a) but is narrower than that of the director as shown in Fig. \ref{fig4}(c) (red-dash line). Furthermore, the photonic sphere-array NA has a degree of polarization-independent feature and is able to redirect both $z$- and $y$-polarized emissions (excluding $x$-polarized one) from an ensemble of atoms or molecules with random dipole orientations. In future work, we will study the performance of the photonic nanoantenna in a lossy environment (background).

In summary, we design a directional and selective photonic NA comprising a single sphere reflector and a sphere chain director. The high directionality originates both from the backward reflection by the Fano resonance and from the forward direction by the dipole resonance. The seamless wavelength selectability is realized by matching the operating wavelength of the reflector with that of the director via tuning the geometrical configurations. Moreover, the smaller loss of photonic elements, compared with the metallic ones, can induce stronger directionality. The photonic NAs may find their applications in photon detection, microscopy, and spontaneous emission management.

The authors acknowledge the support of the grants (Nos. 712010, 711609, and 711511) from the Research Grant Council of the Hong Kong and from The National Natural Science Foundation of China (No. 60931002). This project is also supported in part by a Hong Kong UGC Special Equipment Grant (SEG HKU09) and by the University Grants Council of Hong Kong (No. AoE/P-04/08).

\end{document}